# PERFORMANCE COMPARISON OF VARIOUS ROUTING PROTOCOLS IN DIFFERENT MOBILITY MODELS

Neha Rani[1], Preeti Sharma[2] and Pankaj Sharma[3]

[1]Department of Computer Science, A.B.E.S. Engineering college, Ghaziabad, U.P, India
(nehachandel83@gmail.com)

[2]Department of Computer Science, A.B.E.S. Engineering college, Ghaziabad, U.P, India
(Preeti.sharma860@gmail.com)

[3]Department of Computer Science, Asst. Professor, A.B.E.S. Engineering college, Ghaziabad, U.P, India
(sharma1.pk@gmail.com)

## ABSTRACT

*Mobile Ad hoc Network (MANET) is a infrastructure less network in which two or more devices have wireless communication which can communicate with each other and exchange information without need of any centralized administrator. Each node in the ad hoc network acts as a router, forwarding data packets for other nodes. The main issue is to compare the existing routing protocol and finding the best one. The scope of this study is to test routing performance of three different routing protocols (AODV, OLSR and DSDV) with respect to various mobility models using NS2 simulator. In this paper the parameters used for comparison are packet delivery fraction (PDF), average end to end delay (AEED), normalized routing load (NRL) and throughput.*

## 1. INTRODUCTION

A Mobile Ad-Hoc Network is a collection of mobile nodes with no pre-established infrastructure, self organizing wireless network which forms a temporary network [1].

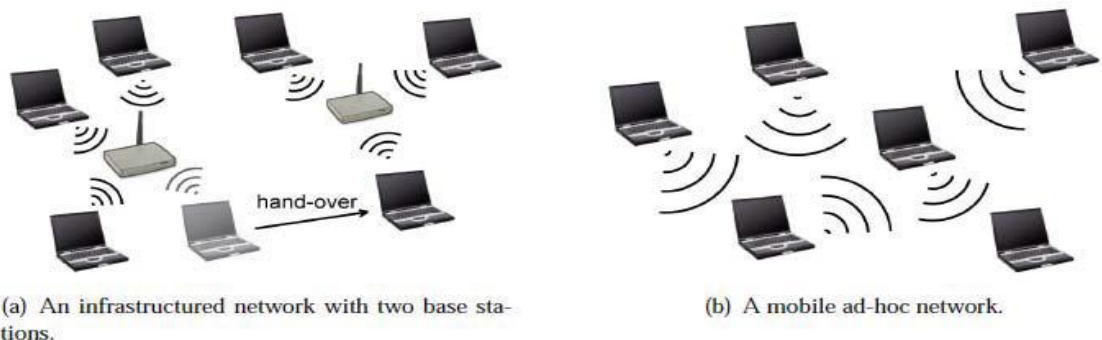

Figure 1: Infrastructured and ad-hoc networks.

Each of the nodes has a wireless interface and communicates each other over either radio or infrared signals. In ad hoc networks [2] all nodes are mobile and can be connected dynamically in an arbitrary manner. One area of research, which has been a focal point of research in Ad hoc networks, is Routing. Generally, Adhoc routing protocols can be classified broadly into two categories, these are Proactive, Reactive. A brief classification of Ad-hoc routing protocols is given in Figure 2.

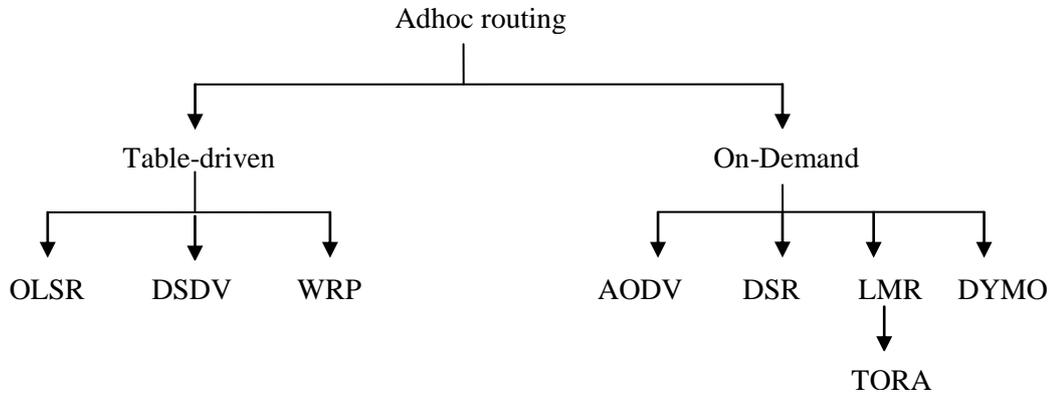

Figure 2: Classification of Routing Protocols in MANET

**Table-Driven routing protocols (Proactive)** In table driven/ proactive routing protocols [4], nodes periodically exchange routing information and attempt to keep up-to-date routing information [5]. Proactive protocols are called so because they have to maintain information about routing prior to its use. Information about routes is maintained in tables called routing tables and when topology changes these tables are updated.

**On Demand routing protocols (Reactive)** In on demand/ reactive routing protocols [6], nodes only try to find a route to a destination when it is actually needed for communication. Proactive protocols are named so, because routing table [7] is not maintained in it. When communication between nodes is required then route is discovered in on-demand manner.

Many routing protocols have been proposed, but just few comparison studies have been performed. Almost all available comparative studies have performed simulations for proactive and reactive protocols by varying number of nodes, network topology, and network density. This paper focuses on varying mobility speed from low to high by keeping other parameters like pause time, number of nodes, mobile connections, simulation duration constant. This study is performed by varying mobility speed and measuring different quantitative metrics in different mobility models i.e. RPGM and RWPM. Reference Point Group Mobility (RPGM) model shows the random motion of mobile nodes associated with a group [3]. Random Waypoint Model (RWPM) assumes that each host is initially placed at random position within the simulation area. As the simulation progresses, each host possess at its current location for a determinable period called the pause time [3].

The remainder of the paper is organized as follows. After presenting the related work in section 2, section 3 presents routing in MANET. Section 4 presents mobility models. Section 5 describes simulation environment. The results of our simulation are analyzed in section 6. Finally, section 7 concludes the paper.

## 2. RELATED WORK

Several researchers have done the quantitative and qualitative analysis of Ad hoc Routing Protocols by means of different performance parameters. Also they have used different simulators for this purpose.

1. Nilesh P.Bopade, Niket N.Mhala performed simulations for comparison of Proactive and Reactive protocols. They have used NS2 Simulator. They have used varying number of mobile nodes but other factors like pause time, speed were not taken into consideration.
2. *J Broch et al.,* performed experimental performance comparison of both Proactive and Reactive routing protocols. In their NS-2 simulation, a network density of 50 nodes with varying pause times and various movement patterns were chosen.
3. *Jorg D.O.* [3] studied the behaviour of different routing protocols for the changes of network topology which resulting from link breaks, node movement, etc. This paper has focussed on performance evaluation by changing number of nodes. But he did not investigate the performance of protocols under high mobility, large number of traffic sources and larger number of nodes in the network which may lead to congestion situations.
4. *Arunkumar B R et al.* Authors perform simulations by using NS-2 simulator. Their studies have shown that reactive protocols perform better than table driven (proactive) protocols.
5. *N Vetrivelan & Dr. A V Reddy* analyzed the performance differentials using varying network density and simulation times. They performed two simulation experiments for 10 & 25 nodes with simulation time up to 100 sec.

## 3. ROUTING IN MANET

This section explains the rouing methodology of various routing protocols taken in comparative study.

### 3.1. Destination sequenced distance vector (DSDV) protocol

One of the examples of proactive protocol is DSDV. This protocol adds a new attribute, sequence number, to each route table entry at each node. Each node maintains a routing table at its own and which helps in packet transmission.

### 3.1.1. Protocol overview and working

For the transmission of packets each node maintains routing table. The routing table maintained by each node also contains the information for the connectivity to different stations in the network. These stations shows all the available destinations and the number of stations (hops) required to reach each destination in the routing table. The routing entry is tagged with a sequence number which is originated by the destination station. Each station transmits and updates its routing table periodically. The packets being broadcasted between stations indicate a list of accessible stations and number of nodes required to reach that particular station. Routing information is advertised periodically by broadcasting or multicasting the packets. In DSDV protocol each mobile station in the network must constantly advertise its routing table to each of its neighbouring stations. As the information in the table may vary frequently, thus the advertisement should be done on the continuous basis so that every node can locate its neighbours in the network. It ensures the shortest number of stations (hops) required from source station to a destination station.

The data broadcast by each node will contain its new sequence number and the following information for each new route:
– The destination address
– The number of hops required to reach the destination and
– The new sequence number, originally stamped by the destination
After receiving the route information, receiving node increments the metric and broadcasts it. Once the mobile nodes receive the route information they broadcast it on an immediate basis. As the mobile nodes change their position within the network results in breaking their links. These broken links may be detected by the layer2 protocol. When there is a broken link in a network, then immediately that metric is assigned an infinity metric determining that there is no hop and the sequence number is updated. Sequence numbers for infinity metrics are odd numbers and the sequence numbers originating from the mobile hosts are defined to be even number. The broadcasting is done in two ways: full dump and incremental dump. Full dump broadcasting will carry all the routing information and requires multiple network protocol data unit (NPDU) while the incremental dump will carry only information that has changed since last full dump and requires only one NPDU to fit in all the information .When an information packet is received from another node, In the first step, it compares the sequence number of the Node with the available sequence number for that entry. If the number of the sequence is larger than the previous one then it will update the routing information with the new sequence number else if the information arrives at node with the same sequence number it looks for the another metric entry and if the number of hops is less than the previous entry the new information is updated at the node (if information is same or metric is more then it will discard the information). While the nodes information is being updated the metric is increased by 1 and the sequence number is also increased by 2. Similarly, if a new node enters the network area, it will announce itself in the network and the nodes in the network update their routing table. During the process of broadcasting, the mobile hosts transmits their routing tables periodically but due to the continuous movements by the hosts in the networks, this will lead to continuous burst of new routes transmissions upon every new sequence number from that destination. The probable solution for this problem is to delay the advertisement of such routes until it shows up a better metric. The Address stored in the routing table corresponds to the layer at which the DSDV protocol is operated [8].

### 3.2. Opitmized link state routing (OLSR) protocol

OLSR is an optimization version of a pure link state protocol. Whenever there is any change in the topology then information is flooded to all nodes. This causes overheads and such overheads are reduced by Multipoint relays (MPR). Two types of control messages are used in OLSR they are topology control and hello messages. There is also Multiple Interface Declaration (MID) messages which are used for informing other host that the announcing host can have multiple OLSR interface addresses [9]. The MID message is broadcasted throughout the entire network only by MPRs. There is also a "Host and Network Association" (HNA) message which provides the external routing information by giving the possibility for routing to the external addresses. Routing in OLSR is described as follows.

#### 3.2.1. Neighbour Sensing

The link in the ad hoc network can be either unidirectional or bidirectional so the host must know this information about the neighbors. The Hello messages are broadcasted periodically for the neighbor sensing. Only nearby neighbour receives hello messages.

#### 3.2.2. Multipoint Relays

Overheads are reduced with the help of MPRs .Instead of pure flooding the OLSR uses MPR to reduce the number of the host which broadcasts the information throughout the network [10].

#### 3.2.3. Multipoint Relays Selection

Proposed algorithm for selecting Multipoint Relay set:
1. All neighbour which want to become MPR are taken.
2. For every neighbour a host degree is calculated. Host degree is actually the number of neighbours whose distance from source is two hops.
3. Then neighbour is added to MPR set .If it is the only neighbour from which is possible to get to the specific two hop neighbour, and then remove the chosen host neighbours from the two hop neighbour set.
4. If there are still some hosts in the two hop neighbour set, then calculate the reach ability of the each one hop neighbour, meaning the number of the two hop neighbours, that are yet uncovered by MPR set.

### 3.2.4. Routing Table Calculations

The host maintains the routing table, the routing table entries have following information: destination address, next address, number of hops to the destination and local interface address.

### 3.3. Ad Hoc On-Demand Distance Vector Routing (AODV)

The Ad Hoc On-Demand Distance Vector routing protocol (AODV) is designed to improve the performance of the Destination-Sequenced Distance Vector routing protocol (DSDV). The main goal of AODV is to broadcast discovery packet when necessary and to distinguish between local connectivity and topology maintenance. In AODV [11] overhead is reduced as number of broadcast is minimized.

### 3.3.1. Path Discovery Process

The process of path discovery starts when a node needs communication with other node by sending route request packet (RREQ) packet [11] which contain broadcast id, source address, destination address, source sequence number, destination sequence number, hop count. When an intermediate node receives RREQ it checks that it had received over bi-directional link. If this has already processed then RREQ packet is discarded. Otherwise, it checks for route entry for destination. The reply is send to source only if the destination sequence number in RREQ is greater than destination sequence number in its route table. A route reply packet (RREP) [11]is send by intermediate node as a response to RREQ packet. As RREP travels back to source all information are updated. Finally, RREP reaches source and route entry is modified.

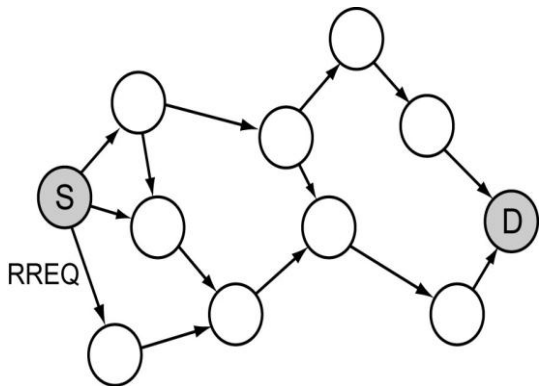

Figure 3: Source node S initiates the path discovery process

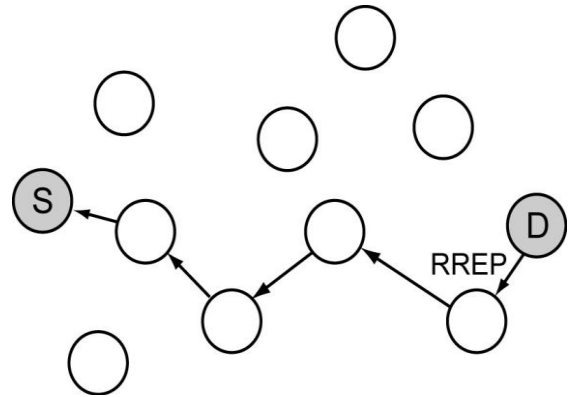

Figure 4: A RREP packet is sent back to the source

### 3.3.2. Maintaining Routes

In AODV [12] each node maintains a routing table with its entries. An active route entry is one in which is in use by active neighbours. Path which is followed by packets from source to destination with active route entries is called an active path. To transmit data from source to destination each time route entry is used.

## 4. MOBILITY MODELS

A model that depicts the movement of mobile nodes, and changes in their velocity and acceleration over time is called Mobility model. Basic parameters related to node movement are number of nodes, mobility speed, pause time, sending rate, number of connections, simulation duration. Mobility models can be classified in to two types group and entity models. In entity models, the motion of mobile nodes are independent from each other, while in group models the movements of nodes are dependent on each other or on some predefined leader node [13].

### 4.1. Reference Point Group Mobility (RPGM)

It is a group mobility model that shows the random motion of mobile nodes. In this model nodes are dependent on some predefined leader node that determines the group motion behaviour.

### 4.2. Random Waypoint Mobility (RWPM)

It is an entity model, in which a node can choose any random velocity and any random destination. The node starts moving towards the selected destination. After reaching the destination, the node stops for a small duration defined by the "Pause Time" parameter and it repeats the complete process again until the simulation process ends.

## 5. SIMULATION ENVIRONMENT

### 5.1. Simulation Model

Here we perform the experiments for the evaluation of the performance of Ad Hoc routing protocol AODV, DSDV, OLSR with varying the mobility speed. We have 30 simulation run in total out of which 15 trace files has been generated for RPGM and RWPM each. We tested all performance metrics in our experiment under varying mobility speed of node (10 to 50m/sec) and while other parameters (nodes = 20, data sending rate = 5 pkts/sec and no. of connections = 10, simulation duration=100sec, pause time=null) are constant.

### 5.2. NS-2 simulator

The network simulations have been done using network simulator NS-2 [14]. The network simulator NS-2 is discrete event simulation software for network simulations. It simulates events such as receiving, sending, dropping and forwarding packets. The ns-allinone-2.34 supports simulation for routing protocols for ad hoc wireless networks such as AODV, DSDV and DSR. NS-2 is written in C++ programming language with Object Tool Common Language. Although NS-2. 34 can be built on different platforms, for this paper, we chose a Linux platform i.e. FEDORA 13, as Linux offers a number of programming development tools that can be used with the simulation process. To run a simulation with NS-2.34, the user must write the OTCL simulation script. The performance parameters are graphically visualized in GRAPH. Moreover, NS-2 also offers a visual representation of the simulated network by tracing nodes events and movements and writing them in a file called as Network animator or NAM file.

### 5.3. Simulation Parameters

This paper considers a network of nodes placing within a 2000m X 2000m area. The performance of AODV, OLSR and DSDV is evaluated by varying the node speed and keeping the other parameters such as number of nodes, transmission rate, pause time, simulation duration constant. Table 1 depicts the simulation parameters used in this evaluation. All performance metric are checked under the varying nodes speed from 10 to 50 m/sec in RPGM and RWPM mobility model.

| SIMULATION PARAMETERS | |
|---|---|
| Simulators | NS-2.34 |
| Protocols | AODV,OLSR ,DSDV |

| SIMULATION PARAMETERS | |
|---|---|
| Simulation duration | 100 sec |
| Simulation area | 2000m X 2000m |
| Pause time | Null |
| Movement model | RPGM,RWPM |
| MAC layer protocol | IEEE 802.11 |
| Traffic type | CBR |
| Data payload | 512 bytes |

Table 1: Simulation table

### 5.4. Performance Metrics

While analyzed the AODV, DSDV and OLSR, protocols, we focused on four performance metrics for evaluation which are Packet Delivery Fraction (PDF), Average End-to-End Delay, Normalized, Routing Load (NRL) and Throughput.

### 5.4.1. Packet delivery fraction

Packet delivery fraction (PDF) is the ratio of number of received data packets successfully at the destinations over the number of data packets sent by the CBR sources.

### 5.4.2. Average End to end delay

It is the average time from the transmission of a data packet at a source node until packet delivery to a destination which includes all possible delays caused by buffering during route discovery process, retransmission delays, queuing at the interface queue, propagation and transfer times of data packets.

### 5.4.3. Normalized Routing Load

The normalized routing load (NRL) it is the ratio of all routing control packets send by all nodes to number of received data packets at the destination nodes.

### 5.4.4. Throughput

It is the average number of messages successfully delivered per unit time number of bits delivered per second.

## 6. SIMULATION RESULTS AND ANALYSIS

The results after simulation are viewed in the form of bar graphs. The performance of AODV, OLSR and DSDV based on the varying the mobility speed is done on parameters like packet delivery fraction, average end-to-end delay, normalized routing load and throughput.

## 6.1. Packet Delivery Fraction

Figure 5 and Figure 6 shows that the group model RPGM is superior compared to entity model i.e. Random Waypoint. This happens because it is a group mobility model and the whole communication process occurs between a few groups. When node speed is 10m/sec, Figure 5 depicts that AODV performed better compared to OLSR and DSDV. But as mobility speed goes up the delivery ratio of routing protocols also went down.

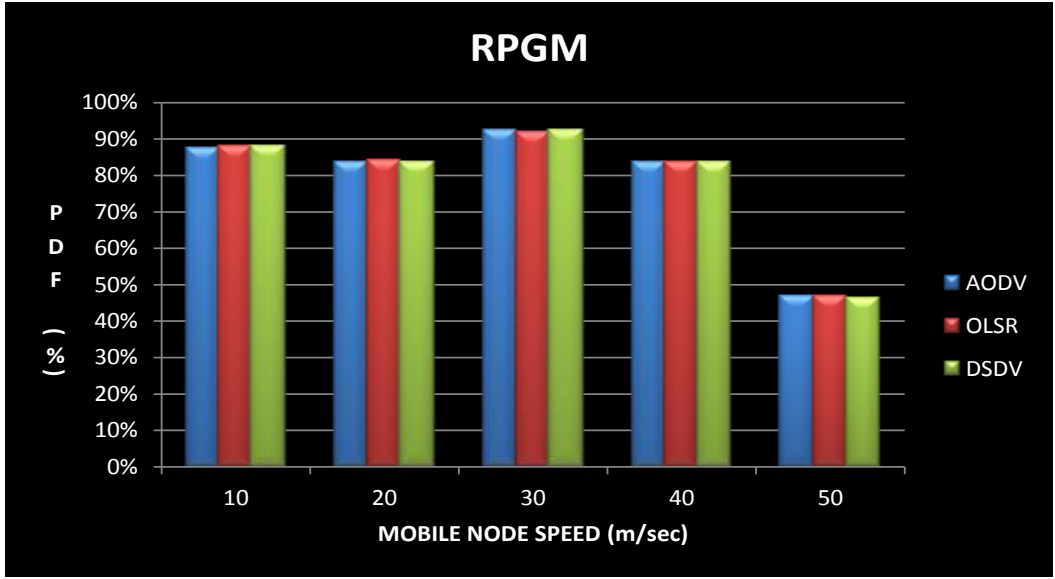

Figure 5: Mobility speed Vs Packet delivery fraction in RPGM Model

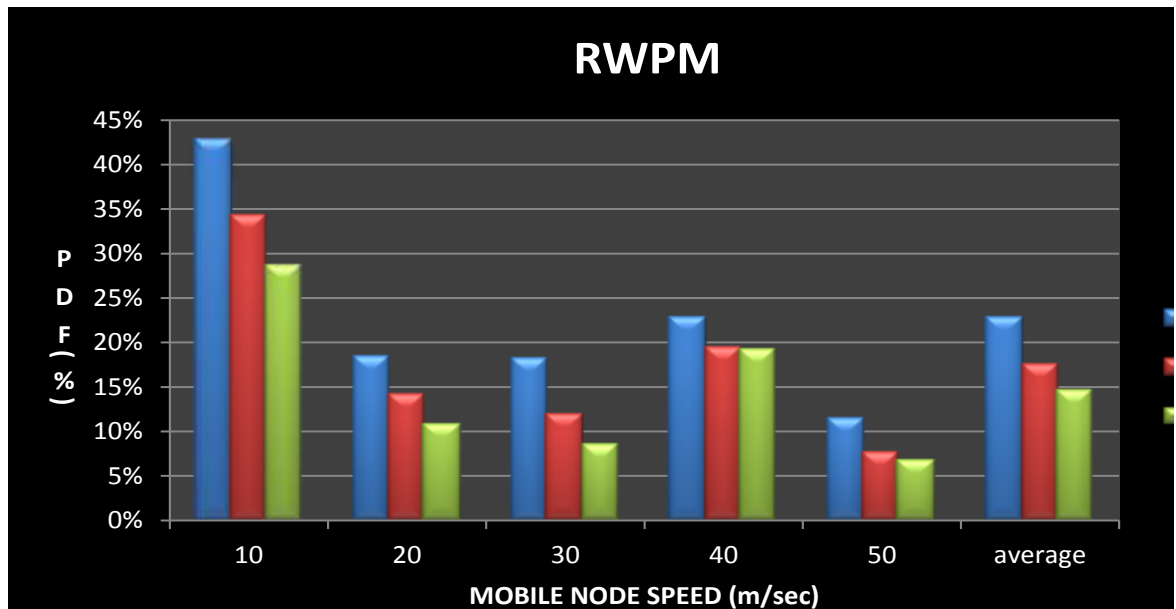

Figure 6: Mobility speed Vs Packet delivery fraction in RWPM Model

## 6.2. Average E2E Delay

This experiment shows that RPGM Model (Figure 8) demonstrates little high average delay for AODV than OLSR. In RWPM, as the mobility speed increases to 50m/sec, the delay of AODV drops down. The performance of OLSR and DSDV is almost equal (an average delay of 5.5ms) .

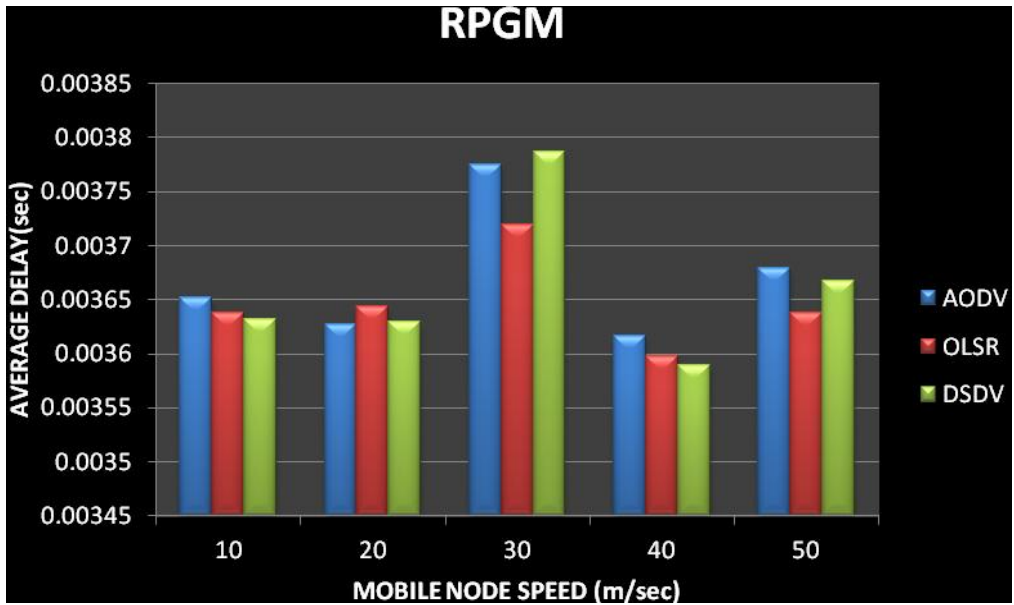

Figure 7: Mobility speed Vs Average delay in RPGM Model

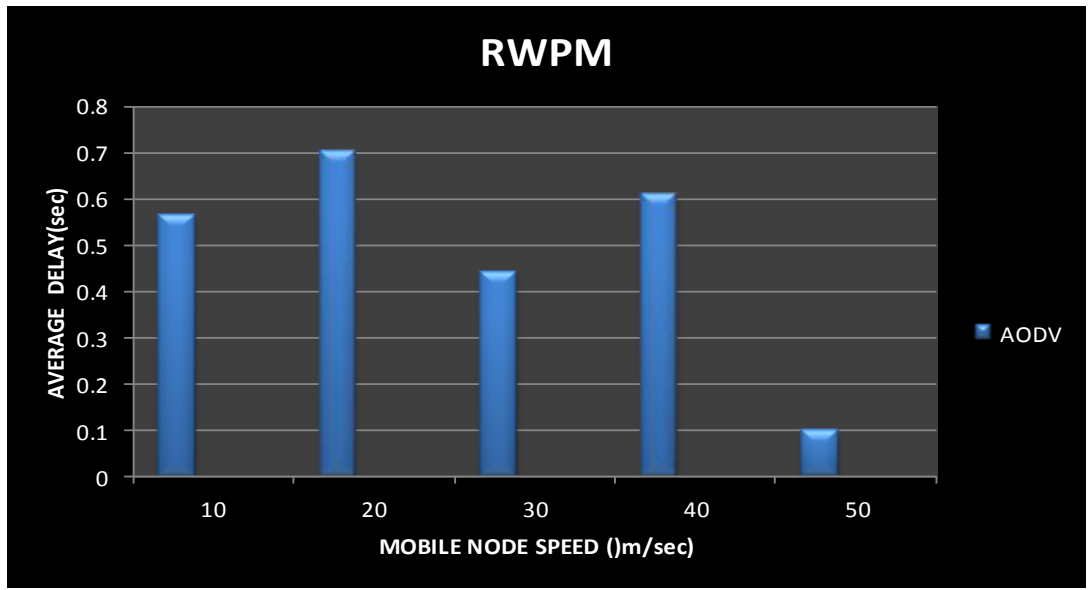

Figure 8: Mobility speed Vs Average delay in RWPM Model

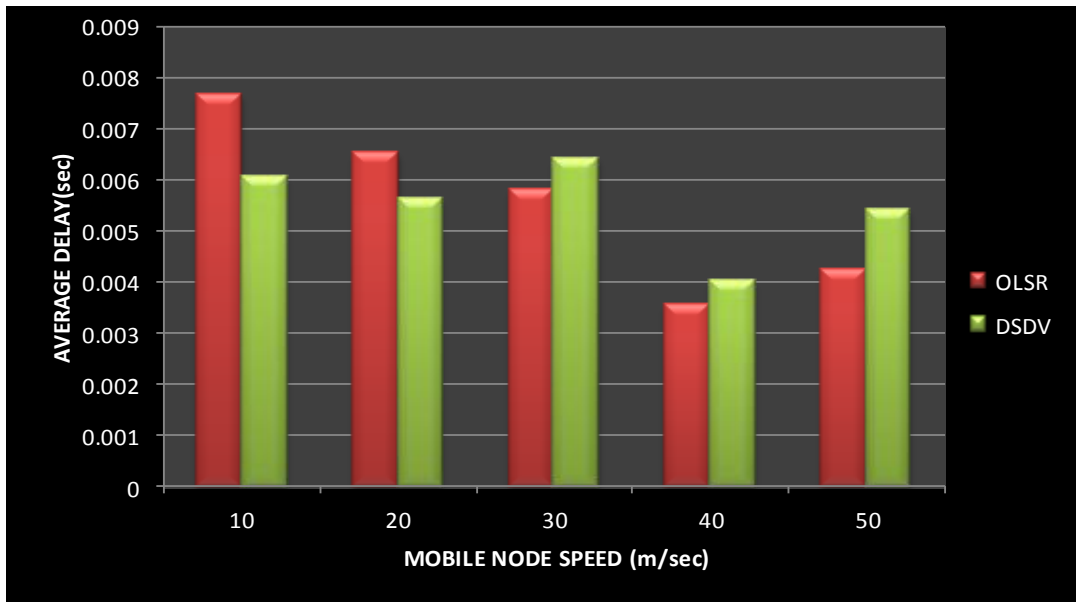

Figure 9: Mobility speed Vs Average delay in RWPM Model

## 6.3. Routing Load

This experiment investigated the routing load of 20 nodes as speed varies from 10 to 20m/sec. Figure 10 and Figure 11 shows routing load of routing protocols in RPGM and RWPM mobility models respectively. Compared with the AODV, OLSR and DSDV, AODV demonstrates the lowest and OLSR shows highest average routing load for all mobility models. In RWPM, the routing load of OLSR is comparatively more. AODV performs better than DSDV and OLSR at the lowest speed level because it is on-demand protocol. This protocol performs best with the RW model.

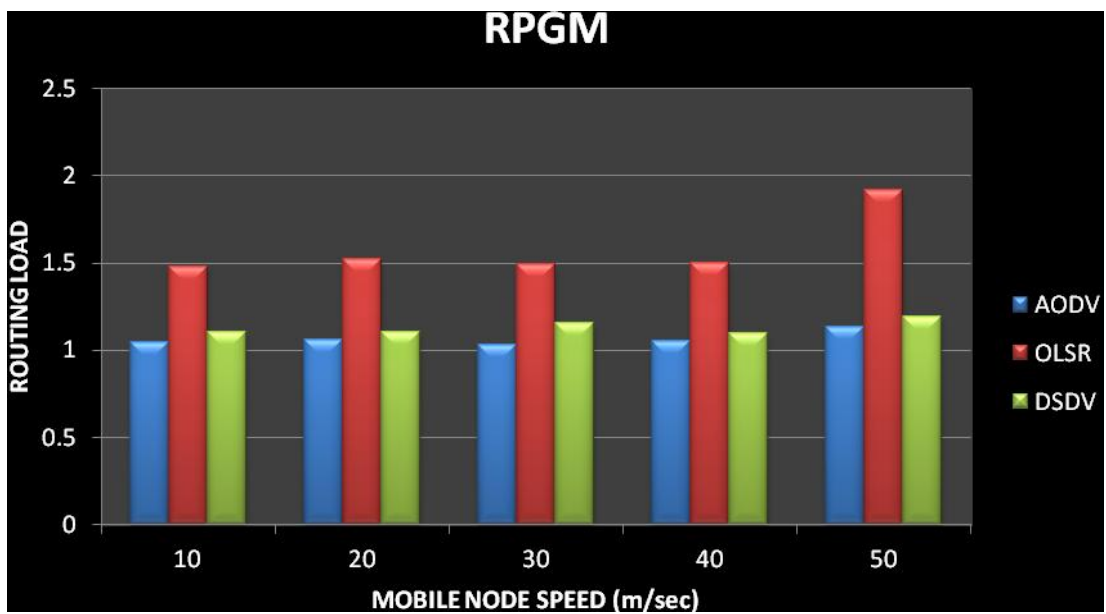

Figure 10: Mobility speed Vs Routing load in RPGM Model

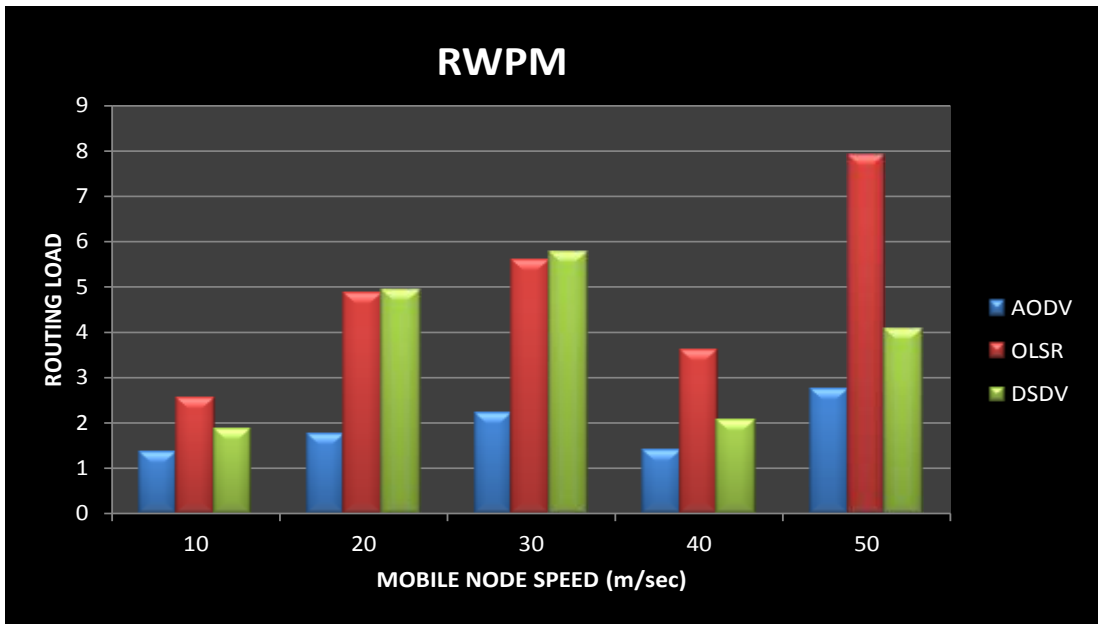

Figure 11: Mobility speed Vs Routing load in RWPM Model

**6.4. MAC Load**

In this experiment, Figure 12 depicts that AODV, OLSR and DSDV all shows nearly same performance in RPGM. However Figure 13 demonstrates that the MAC load of AODV is approx. 8.573(low) and for DSDV is 19.614(high) in RWPM.

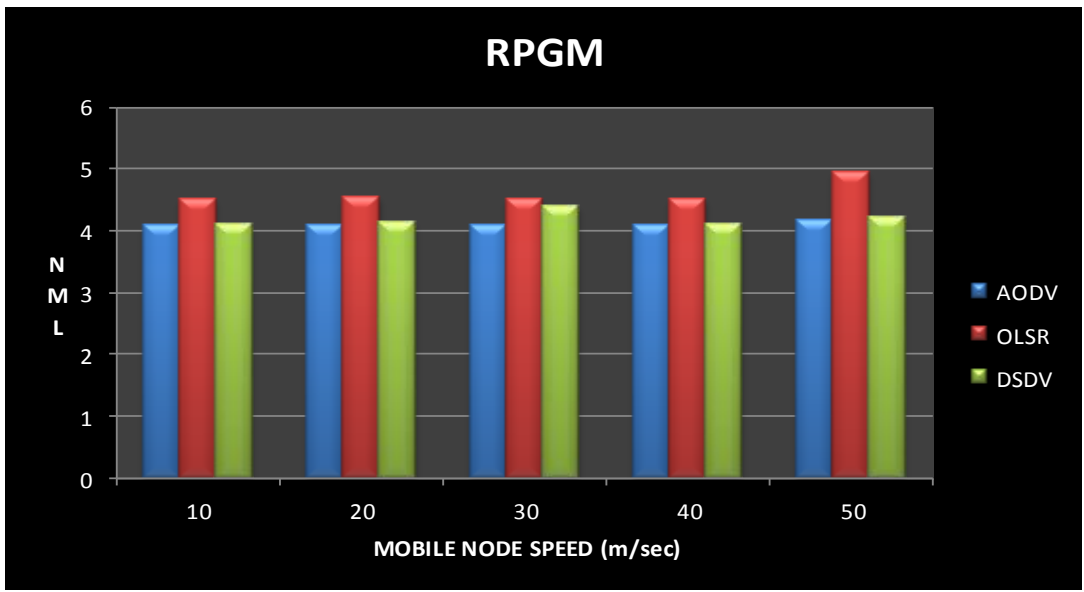

Figure 12: Mobility speed Vs Normalized MAC load in RPGM Model

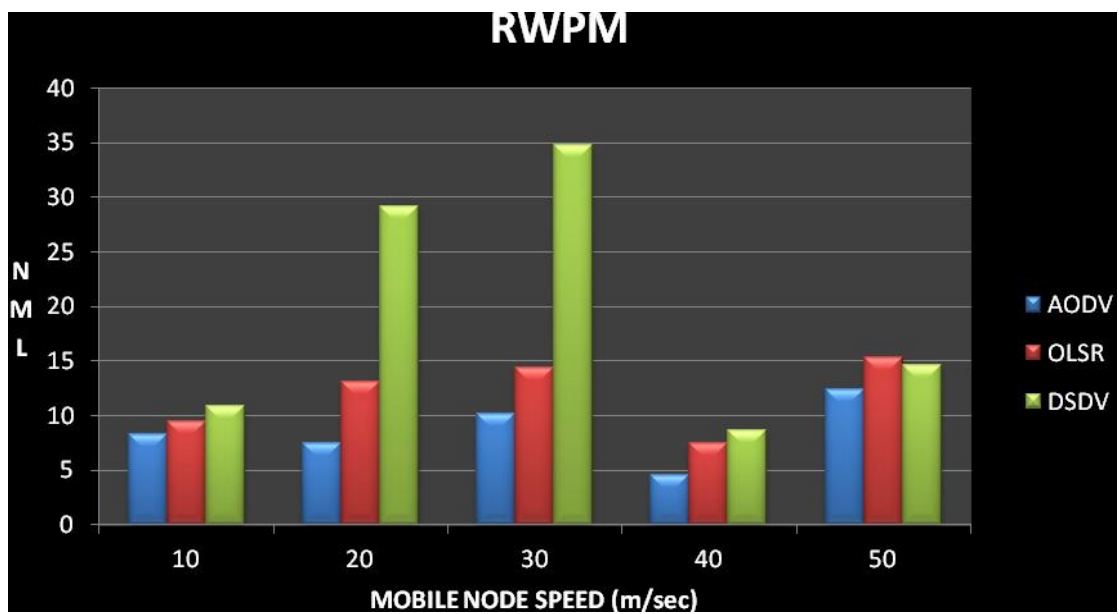

Figure 13: Mobility speed Vs Normalized MAC load in RWPM Model.

## 7. CONCLUSION

This paper studied performance of the three widely used MANET routing protocols (AODV, OLSR and DSDV) with respect to group (RPGM) and entity (RWPM) mobility models. We have developed a set of simulation scripts for the NS2 simulation environment merged with the BonnMotion scenario generation tools. Simulation results have indicated that the relative ranking of routing protocols may vary depending on mobility model. The relative ranking also depends on the node speed.

Analysis on RPGM model shows that message delivery rate (packet delivery ratio) of AODV, OLSR, and DSDV was almost same but as the mobility increases the message Delivery Rate goes on decreasing. Hence it is concluded that the PDF of AODV is comparatively higher. In case of Average End to End delay (AEED), the OLSR shows least delay as compared to AODV and DSDV. The study shows that the AODV demonstrates lowest routing load and OLSR shows its highest values. For performance metric, MAC load OLSR shows maximum values. Hence it can be concluded that In RPGM model, AODV shows good performance. In RWPM the AODV showed maximum packet delivery ratio. In case of average delay as mobility increases the OLSR shows low delays. Routing load and MAC load of AODV is least as compared to OLSR and DSDV routing protocols.

RPGM model is suitable for military battlefield, disaster management and other rescue operations areas, where high message delivery rate with low routing load, low MAC load and delay is required, Hence considering simulation results AODV is best suited protocol for this model. Future work should be focused to extending set of the experiments by taking into consideration the simulation parameters, different propagation models and MAC protocols.